\documentclass[10pt,a4paper,twocolumn,tightenlines,amsmath,amssymb,nofootinbib,superscriptaddress,showpacs]{revtex4-1}

\usepackage[UKenglish]{babel}

\usepackage{natbib}
\usepackage{hyperref}
\usepackage{cleveref}
\crefname{figure}{figure}{figures}

\usepackage{amsmath,amssymb,amsthm}
\usepackage{bm}
\usepackage{bbold}
\usepackage{array}

\usepackage{latexsym,epsfig,epstopdf}
\usepackage{graphicx}
\usepackage[font=small]{caption}
\usepackage{subcaption}
\newcommand{\caphead}[1]{{\bf #1}}

\makeatletter
\DeclareRobustCommand{\cev}[1]{%
  \mathpalette\do@cev{#1}%
}
\newcommand{\do@cev}[2]{%
  \fix@cev{#1}{+}%
  \reflectbox{$\m@th#1\vec{\reflectbox{$\fix@cev{#1}{-}\m@th#1#2\fix@cev{#1}{+}$}}$}%
  \fix@cev{#1}{-}%
}
\newcommand{\fix@cev}[2]{%
  \ifx#1\displaystyle
    \mkern#23mu
  \else
    \ifx#1\textstyle
      \mkern#23mu
    \else
      \ifx#1\scriptstyle
        \mkern#22mu
      \else
        \mkern#22mu
      \fi
    \fi
  \fi
}

\newcommand{\past}[1]{\cev{#1}}
\newcommand{\future}[1]{\vec{#1}}

\makeatletter
\newcommand*{\balancecolsandclearpage}{%
  \close@column@grid
  \clearpage
  \twocolumngrid
}
\makeatother

\newcommand{\ket}[1]{\left | #1 \right\rangle}

\newcommand{\braket}[2]{\left\langle #1|#2\right\rangle}
\newcommand{\ketbra}[2]{|#1\left\rangle\right\langle #2 |}

\newcommand{\Fo}[1]{\mathcal{F}\!\left( #1 \right)}

\newcommand{\conv}[0]{\ast}

\DeclareMathOperator{\tr}{Tr}

\DeclareMathOperator{\sinc}{sinc}

\providecommand{\Pr}{}
\renewcommand{\Pr}[1]{\mathrm{P}\!\left(#1\right)}
\newcommand{\cPr}[2]{\mathrm{P}\!\left(#1 \,|\, #2\right)}

\newcommand{\Ent}[1]{{H}\hspace{-0.25em}\left(#1\right)}

\newcommand{\cPrline}[2]{\mathrm{P}(#1 \,|\, #2)}
\newcommand{\Prline}[1]{\mathrm{P}(#1)}

\newcommand{\inlineheading}[1]{\textbf{{#1}}}
\newcommand{\inlinesubheading}[1]{\textit{#1}}

\usepackage{color}
\usepackage[dvipsnames]{xcolor}
\usepackage[normalem]{ulem}
\usepackage{xspace}

\newcommand{\CQT}{Centre for Quantum~Technologies, National~University~of~Singapore, 3 Science Drive 2, 117543, Singapore}
\newcommand{\Tsing}{Center for Quantum~Information, Institute~for~Interdisciplinary~Information~Sciences, \\
Tsinghua~University, Beijing, 100084, China}
\newcommand{\Oxf}{Atomic~and~Laser~Physics, University~of~Oxford, Clarendon~Laboratory, Parks~Road, Oxford, OX1 3PU, United Kingdom.}
\newcommand{\NTUPhys}{School of Physical and Mathematical Sciences, Nanyang Technological University, 639673, Singapore}
\newcommand{\NTUComplex}{Complexity Institute, Nanyang Technological University, 639673, Singapore.}

\begin{document}


\title{Provably unbounded memory advantage in stochastic simulation\\ using quantum mechanics.
}

\author{Andrew J. P. Garner}
\email{ajpgarner@nus.edu.sg}
\affiliation{\CQT}
\affiliation{\Tsing}

\author{Qing Liu}
\affiliation{\NTUPhys}

\author{Jayne Thompson}
\affiliation{\CQT}

\author{Vlatko Vedral}
\address{\Oxf}
\affiliation{\CQT}
\address{Department of Physics, National University of Singapore, 2 Science Drive 3, Singapore 117542}
\address{\Tsing}

\author{Mile Gu}
\email{gumile@ntu.edu.sg}
\address{\NTUPhys}
\address{\NTUComplex}
\affiliation{\CQT}

\date{\today}



\begin{abstract}
Simulating the stochastic evolution of real quantities on a digital computer requires a trade-off between the precision to which these quantities are approximated, and the memory required to store them.
The statistical accuracy of the simulation is thus generally limited by the internal memory available to the simulator.
Here, using tools from computational mechanics, we show that quantum processors with a fixed finite memory can simulate stochastic processes of real variables to arbitrarily high precision.
This demonstrates a provable, unbounded memory advantage that a quantum simulator can exhibit over its best possible classical counterpart.
\end{abstract}

\maketitle


Many macroscopic processes we wish to simulate involve the dynamics of real numbers.
The dynamical properties we wish to track (e.g.\ the position of an object) can take on almost any number,
 seemingly without noticeable quantization until one goes down to the Planck scale.
The simulation of such processes necessitates compromise between the resources allocated and the precision with which we track such properties.
Clever implementations to this problem, such as the floating point format~\cite{IEEEfloat}, form the heart of modern computing technology --
 but all subscribe to the same trade-off: treating a quantity with higher precision requires the allocation of more memory. To perfectly replicate the future statistics of a continuous variable dynamical system exactly would inevitably require unbounded memory.

The advent of quantum technology, however, opens new possibilities. 
Not only has this technology shown great potential in solving problems many consider classically intractable~\cite{Deutsch85,DeutschJ92,Grover96,Shor97,CleveEMM98}, 
 it has demonstrated the capability to greatly reduce the amount of information one needs to send in certain tasks requiring communication between distributed parties~\cite{vanDam00,deWolf01,Brassard03}.
Could the memory required by a quantum machine that simulates dynamical processes likewise scale much more favourably with precision?

Here, we consider the simulation of a class of stochastic systems involving the dynamics of parameters that take on real numbers.
Classical simulation of such processes digitally involves `coarse-graining': the parameter at each point in time is approximated to $n$ bits of precision at some memory cost that scales linearly with $n$.
We construct quantum simulators the exhibit \emph{unbounded} advantage.
The quantum simulator can exactly replicate the statistics of a $n$ bit classical simulator for arbitrarily large $n$ using a bounded amount of memory. Thus, quantum simulators can side-step the precision-memory tradeoff -- finite quantum memory can simulate such processes to arbitrary fixed precision.

This unbounded divergence has practical and foundational consequences.
Practically, it suggests that quantum processors may be increasingly advantageous as we wish to simulate ever more memory-intensive systems, such as those arising from big data sets.
Foundationally, the minimal memory required to simulate a process is a well-established measure of structure, known as {\em statistical complexity}~\cite{Grassberger86, CrutchfieldY89, ShaliziC01, CrutchfieldEM09, Crutchfield11, HaslingerKS10, ShaliziSH04, MarquesdaSilva1997, ClarkeFW03, ParkWLYJM07, LiYK08, LuB12}.
Our work suggests that there are certain processes which grow unboundedly in statistical complexity, but yet remain simple to an observer with quantum capabilities.

\begin{figure}[b]
\includegraphics[width=0.3\textwidth]{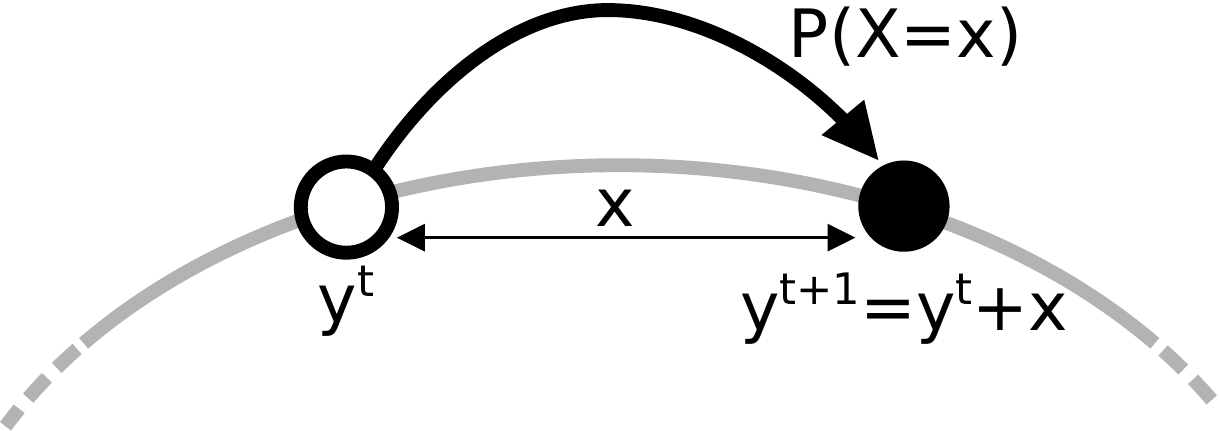}
\caption{
\label{fig:CyclicWalk}
\caphead{Cyclic random walk.} At each time step, the system stochastically hops from state $y^t \in [0,1)$ to $y^{t+1} = \mathrm{frac}[y^t + x]$.
As $x$ is chosen according to the real random variable $X$, the current value of the system is itself described by a sequence of real random variables $\{Y^t\}_{t\in\mathbb{Z}}$ that satisfy $Y^{t+1}=\mathrm{frac}[Y^t + X]$.
}
\end{figure}

%
\vspace{0.5em}
\inlineheading{Cyclic random walks.}
Consider a small bead located on a circular ring of circumference 1 (as per \cref{fig:CyclicWalk}).
Its position can always be described by some real number $y \in [0,1)$.
At each discrete time $t\in\mathbb{Z}$, the bead's position is stochastically perturbed. 
This perturbation is described by a real random variable $X$ that is governed by a continuous probability density function $P(X)$, such that
\begin{equation}
Y^{t+1} = \mathrm{frac}[Y^t + X],
\end{equation}
where $Y^{t}$ represents the random variable that governs the location of the bead at time $t$, 
and $\mathrm{frac}[y] = y - \lfloor y \rfloor \in [0,1)$ denotes the fractional part of $y$,
 such that positions differing only by whole rotations around the ring are equivalent.
We refer to $P(X)$ as the {\em shift function}, and assume the process is stationary, in the sense that $P(X)$ has no explicit dependence on $t$,
 and rotationally symmetric such that $X$ has no dependence on the current value of $Y^t$.
This same formalism describes a diverse range of systems undergoing \emph{cyclic random walks}, such as the azimuthal motion of gas molecules diffusing in an annular tube,
 or the position of a single electron travelling through an electric circuit with constant resistance.

We capture the dynamics of $Y$ formally using the framework for describing stochastic processes.
In general, a stochastic process $\mathcal{P}$ is characterized by a bi-infinite sequence of random variables $\{Y^t\}_t$, that governs its value at each discrete time $t\in\mathbb{Z}$.
For convenience, we often segregate past and future values, such that $\past{Y} = \ldots Y^{-1}Y^{0}$ and $\future{Y} = Y^{1}Y^{2}\ldots$ respectively govern the values in the past and future with respect to time $t=0$.
The cyclic random walk above is then entirely captured by the joint probability distribution $\Prline{\past{Y}, \future{Y}}$ such that for any instance of the process with past values $\past{y}$, future values $\future{y}$ will be observed with probability $\cPrline{\future{Y}=\future{y}}{\past{Y} = \past{y}}$.

\enlargethispage{2\baselineskip}
Here, we consider the simulations of the above process to ever increasing precision.
We adopt {a natural technique of discretizing a continuous process},
 by introducing a family of stochastic processes $\{\mathcal{P}_n\}$ that describe discrete approximations of this process,
 where in each the position of bead is represented to $n$ bits of precision by a $n$-digit binary number.
This is done by limiting $y$ to a discrete set of $N = 2^n$ equally--spaced values, $y_j = j/N$ (for $j=0$ to $N-1$).
At each time-step, the probability that a bead in discrete location $y_j$ transitions to $y_k$,
 is given by the probability $p_{jk}$ that a bead initially at $y_j$ will transition to any value of $y$ whose $n$ bit binary representation is $y_k$.
That is
\begin{equation}
p_{kj} = \cPr{Y^{t+1} = y \in \mathcal{I}_k }{ Y^{t} = y_j}
\end{equation}
where $\mathcal{I}_k = \{y:  |y - y_k|  < \frac{1}{2N} \}$ represents the interval on the ring that is `rounded to' $y_k$.
This results in a Markovian stochastic process that emits a symbol from the finite alphabet $\{y_k\}$ at each time-step,
 whose dynamics are governed by the stochastic matrix with elements $p_{jk}$.
As $n \rightarrow \infty$, the statistics of $\mathcal{P}_n$ approach that of $\mathcal{P}$;
 at the potential cost of tracking more information\footnote{An alternative discretization is to calculate the transition probabilities by assuming the initial value of $y^{t}$ is uniformly distributed in $\mathcal{I}_j$.
This yields asymptotically identical statistics as $N\to\infty$, and does not change the results of this article.}.

\pagebreak
\inlineheading{Classical simulation costs scale with precision.}
We can formally describe simulators using the tools of computational mechanics~\cite{CrutchfieldY89,ShaliziC01,CrutchfieldEM09,Crutchfield11}.
A simulator of a process is a device whose future output behaviour conditioned on any particular past should be statistically indistinguishable to the process itself.
Specifically, let the state of the simulator at each time be $s^{t}$, such that at the subsequent time-step it can output $y^{t+1}$ and transition to state $s^{t+1}$.
For this device to be a statistically faithful simulator of a process $\Prline{\past{Y},\future{Y}}$, we require that:
\begin{enumerate}
\item For each specific past $\past{y}$ at each time $t$, we can deterministically configure the device using a function $f$ into some state $s = f(\past{y})$, such that it will produce future outputs $\future{y}$ with probability $\cPrline{\future{Y} = \future{y}}{\past{Y} = \past{y}}$.
\item If a simulator is in state $s^t = f(\past{y})$ at time $t$, and outputs $y^t$ in the subsequent time-step, its internal state must then transition to $s^{t+1} = f(\past{y} y^t)$.
\end{enumerate}

The first condition ensures the simulator can be initialized to simulate desired conditional future statistics;
 the second that a correctly initialized simulator continues to exhibit statistically correct statistics at every time-step.
The memory cost of the simulator corresponds to the storage requirements of this internal state.
This cost is bounded from below by the information entropy of the random variable $S := f(Y)$.
In the asymptotic limit of many independent identically distributed copies of the simulator, this bound is tight as the ensemble of states may be compressed (such as by Shannon's noiseless encoding theorem~\cite{Shannon48}, or Schumacher compression~\cite{Schumacher95,Winter99}).
Physically a simulator can be viewed as a communication \mbox{channel} in time: it represents the exact object Alice must give to Bob at each time-step that captures sufficient past information for Bob to replicate the processes conditional future behaviour.
$f$ is known as the {\em encoding function}, which describes how the past is encoded within the \mbox{channel}.

This memory cost of the provably-optimal classical simulator -- known as the \emph{statistical complexity} $C_\mu$ -- is extensively studied in complexity science~\cite{CrutchfieldY89}.
This value captures the absolute minimum memory any classical simulator of a process must store, and thus is a prominent quantifier of a process's structure and complexity\footnote{
The statistical complexity is distinct from algorithmic information (Kolmogorov--Chaitin complexity). 
Statistical complexity is, as the name would imply, intrinsically statistical -- concerned with the replication of the statistical behaviour of a process; whereas algorithmic information relates to the compressibility of an exact string~\cite{Ladyman13}.
} (e.g.\ \cite{Crutchfield11,MarquesdaSilva1997,ClarkeFW03,ShaliziSH04,ParkWLYJM07,LiYK08,HaslingerKS10,LuB12}).
Such an optimal simulator can be explicitly constructed, and corresponds to the simulator that stores in its internal memory the {\em causal states} of the process~\cite{CrutchfieldY89,ShaliziC01}: defined by an encoding function $f$ such that $f(\past{y}) = f(\past{y}')$ if and only if $\cPrline{\future{Y}}{\past{Y}=\past{y}} = \cPrline{\future{Y}}{\past{Y}=\past{y'}}$ (i.e. the conditional futures of $\past{y}$ and $\past{y}'$ coincide).

In our cyclic random walks, each $\mathcal{P}_n$ is a first-order Markov process: the statistics of future outcomes depend only on the most recent value of $Y^t$.
When this example is discretized, the causal states are thus typically in one-to-one correspondence with the $2^n$ discrete values that $Y$ can take%
\footnote{There are exceptions, such as when $P(x)\!=\!1$ for $x\in[0,1)$, and the system jumps to a completely random point at each time-step; here there is only one causal state for all $N$, because the current position no longer affects the future outcomes at all.}.
That is, $\mathcal{P}_n$ has $2^n$ causal states, labelled $\{s_j\}_{j=0}^{2^n-1}$, where $s_j$ corresponds to the set of pasts ending in $Y^0 = y_j$. 
When the simulator has been running for a sufficiently long time,
 the probability distribution over the internal memory converges on $\Prline{S\!=\!s_i} = \frac{1}{N}$ for each $i$ -- its steady state, in which all causal states occur with equiprobability. 
Thus, the classical statistical complexity
\begin{equation}
C_\mu = n,
\end{equation}
scales linearly with the precision.

\inlineheading{Quantum simulators are memory--efficient.}
It has recently been shown that quantum processors have the capability to simulate stochastic processes with less memory than is classically possible~\cite{GuWRV12,SuenTGVG15,MahoneyAC16,PalssonGHWP16,RiechersMAC16}.
Here, we construct an explicit quantum simulator for the cyclic random walk. 
Instead of storing each causal state $s_i$ directly, our quantum simulator stores a corresponding quantum state
\begin{equation}
\label{eq:QStates}
\ket{S_j} = \sum_{k=0}^{N-1} \sqrt{p_{kj}}\ket{k},
\end{equation}
where $\{\ket{k}\}$ forms an orthonormal basis.

\begin{figure}[bht]
\includegraphics[width=0.4\textwidth]{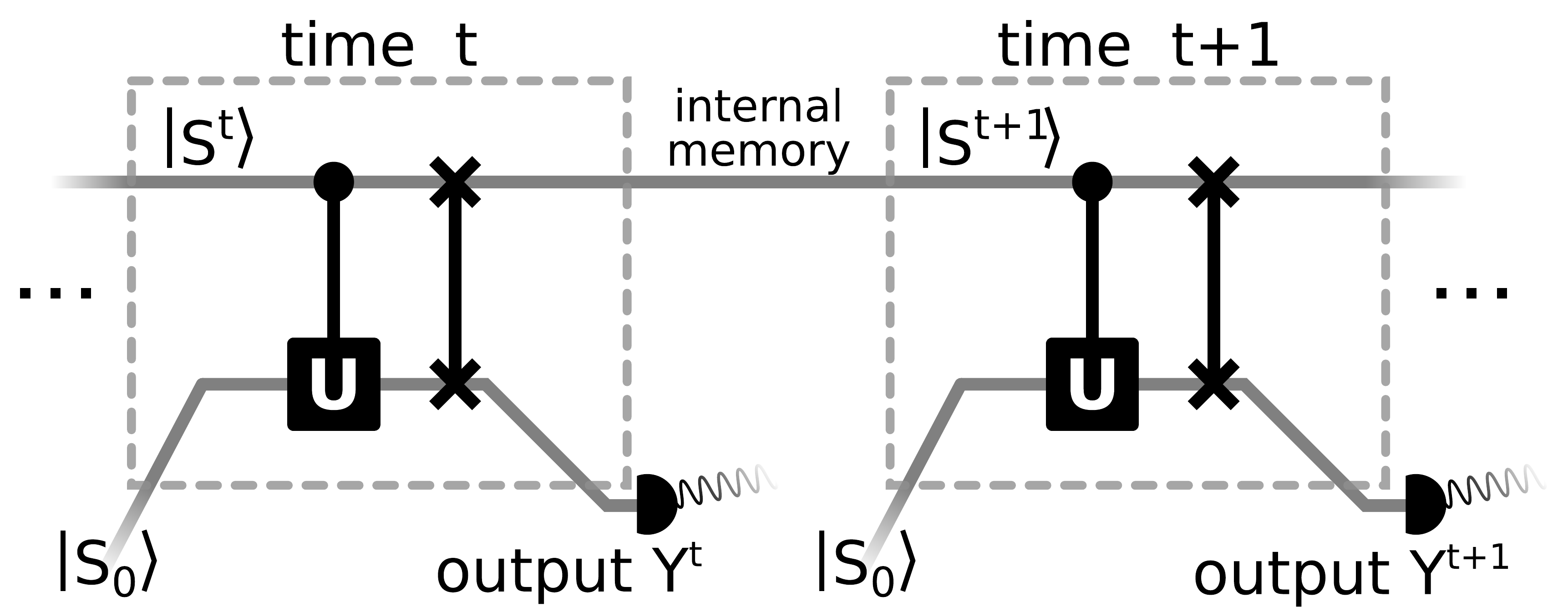}
\caption{
\label{fig:Circuit}
\caphead{Circuit for memory-efficient quantum simulator.}
The above circuit samples $\cPrline{\future{Y}}{\past{y}}$ when supplied with the appropriate quantum state $\ket{S^t}$ that encodes the past.
At $t = 0$, an ancillary system, initialized in state $\ket{S_0}$, is fed into the simulator.
A controlled unitary is then enacted such that $U: \ket{j}\ket{S_0} \rightarrow \ket{j}\ket{S_j}$ for each $j$.
The state of the ancillary system and memory are then coherently swapped, and the ancillary system is then emitted as output.
Measurement of the ancillary system then correct samples $\future{Y}^1$.
Iteration of this procedure then generates output behaviour statistical identical to that of the original process.
}
\end{figure}

The stationary state of the quantum simulator is then given by the \emph{quantum ensemble state} $\rho = \frac{1}{N} \sum_j \ketbra{S_j}{S_j}$ (as all quantum states occur with equiprobability).
Thus the memory required to store these states is given by the von Neumann entropy given $H_Q := -\tr\left(\rho\log\rho\right) = -\sum_k \lambda_k \log \lambda_k$, where $\lambda_k$ are the eigenvalues of $\rho$.
The key improvement here is that $\{\ket{S_j}\}$ are not in general mutually orthogonal, and thus $H_Q$ is generally less than $C_\mu$.
Nevertheless a quantum circuit (outlined in figure~\ref{fig:Circuit} -- with details in the {\em Technical Appendix}) acting on these quantum states will produce statistically identical outputs to the classical simulator.

The von Neumann entropy of a quantum state is equal to the Shannon entropy of the outcome statistics of a projective measurement on that state, minimized over all choices of projective measurement.
This minimization corresponds to a measurement in the basis in which the state's density matrix is diagonal.
A classical probability distribution maps onto a mixed quantum state, diagonal in a fixed basis.
As such, the stationary state of the classical simulator can be assigned a quantum state, whose von Neumann entropy is exactly that distribution's Shannon entropy.
This allows us to compare the entropic cost of the classical and quantum machines' memories on an equal footing.

\inlineheading{Unbounded advantage of quantum memory.}
\mbox{We now come} to the main claim of our paper:
there are stochastic processes that can be simulated to infinite precision using a finite amount of quantum memory.

\begin{figure*}[tb]
\begin{centering}
\begin{subfigure}[t]{0.4\textwidth}
\includegraphics[width=\textwidth]{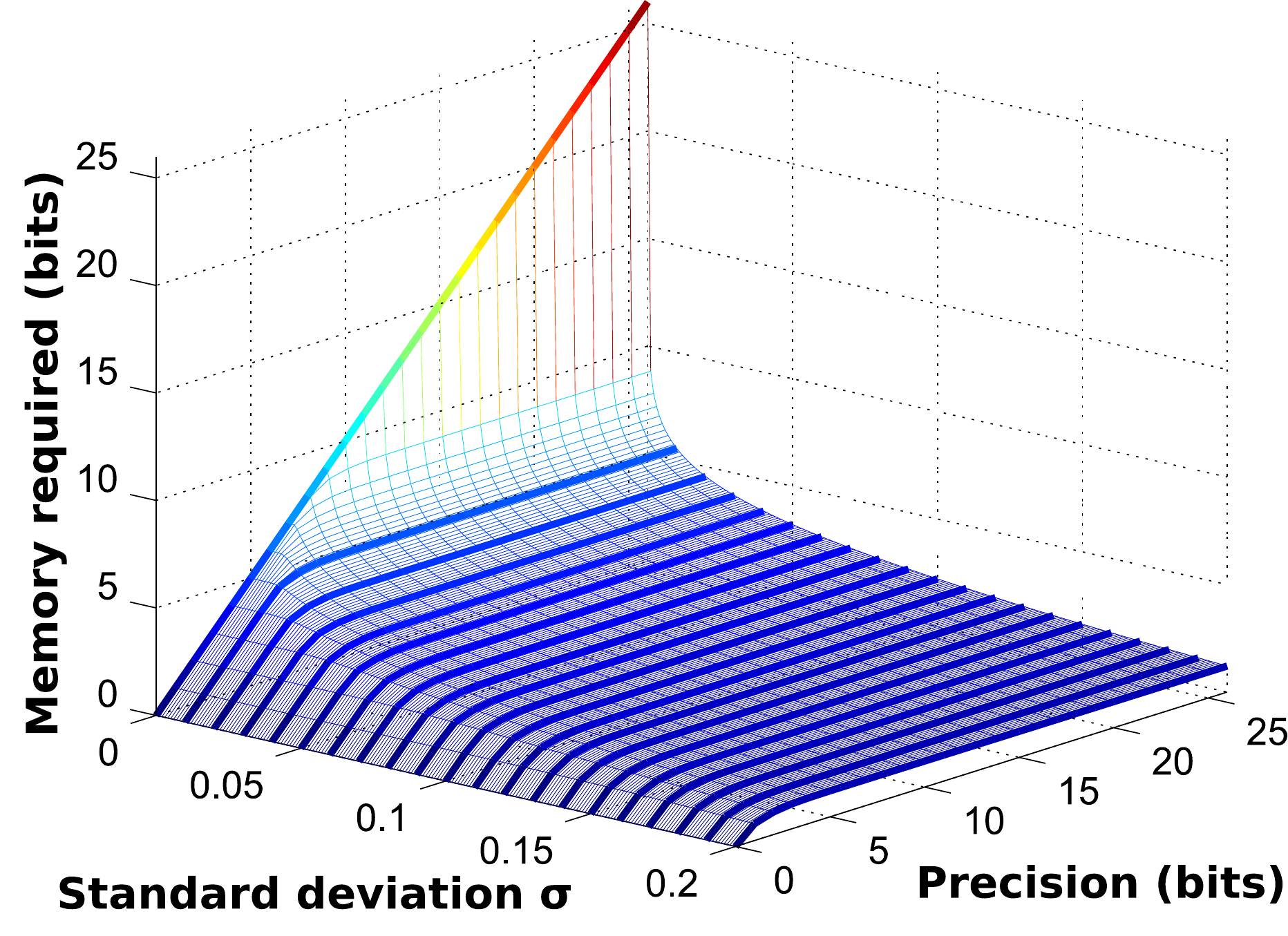}
\subcaption{\label{fig:GPlot} Quantum memory cost for process with Gaussian noise
$P(x) = \dfrac{1}{\sigma\sqrt{2\pi}} \exp\left(-\dfrac{x^2}{2\sigma^2}\right)$.}
\end{subfigure}
\hspace{1em}
\begin{subfigure}[t]{0.4\textwidth}
\includegraphics[width=\textwidth]{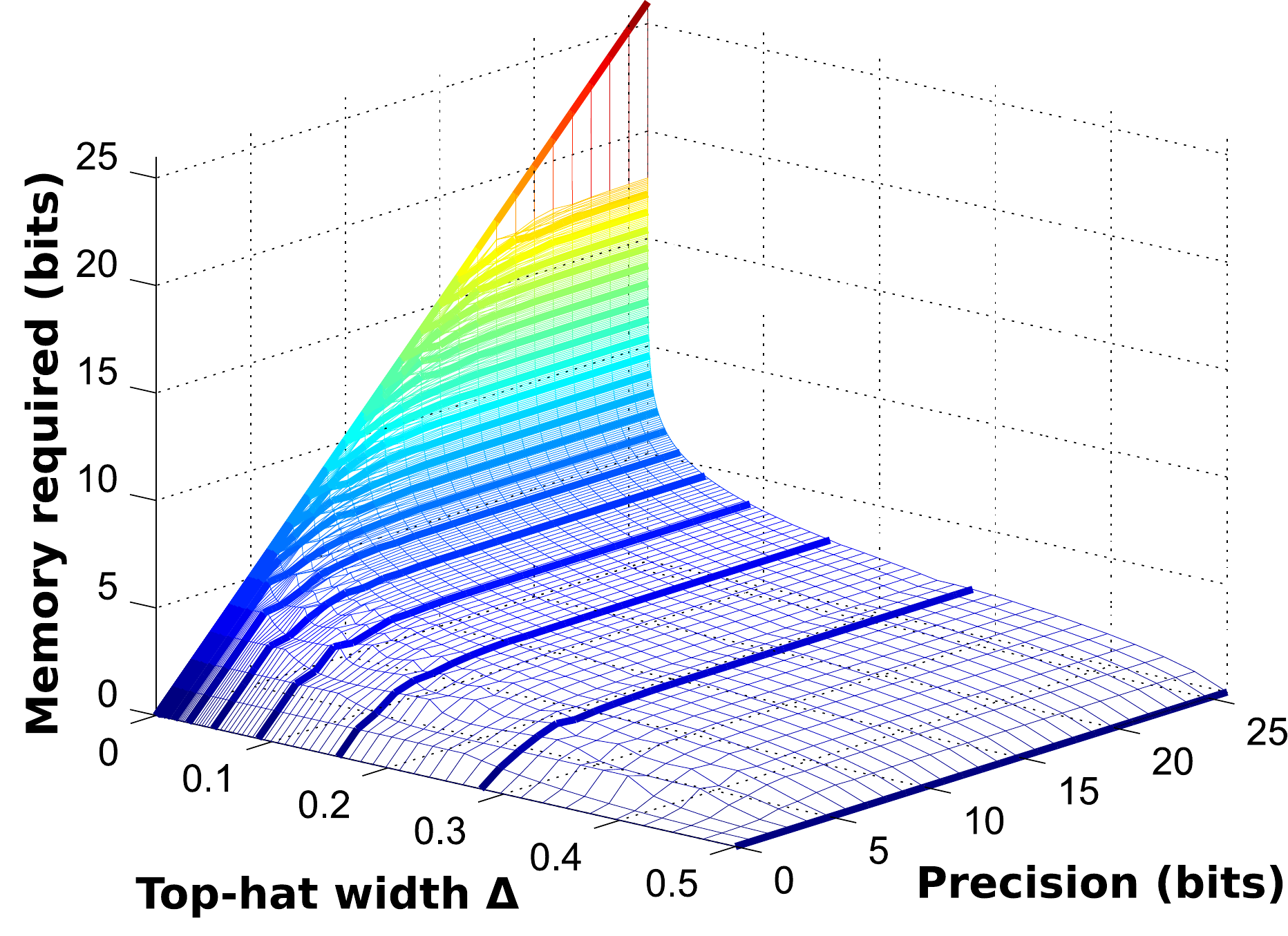}
\subcaption{
\label{fig:THPlot}
Quantum memory cost for process with uniform white noise
$P(x) = \dfrac{1}{2\Delta}$ when $x\in[-\Delta,\Delta]$ and $P(x) = 0$ elsewhere.}
\end{subfigure}
\end{centering}\\
\vspace*{0em}
\begin{centering}
\begin{subfigure}[t]{0.4\textwidth}
\includegraphics[width=0.95\textwidth]{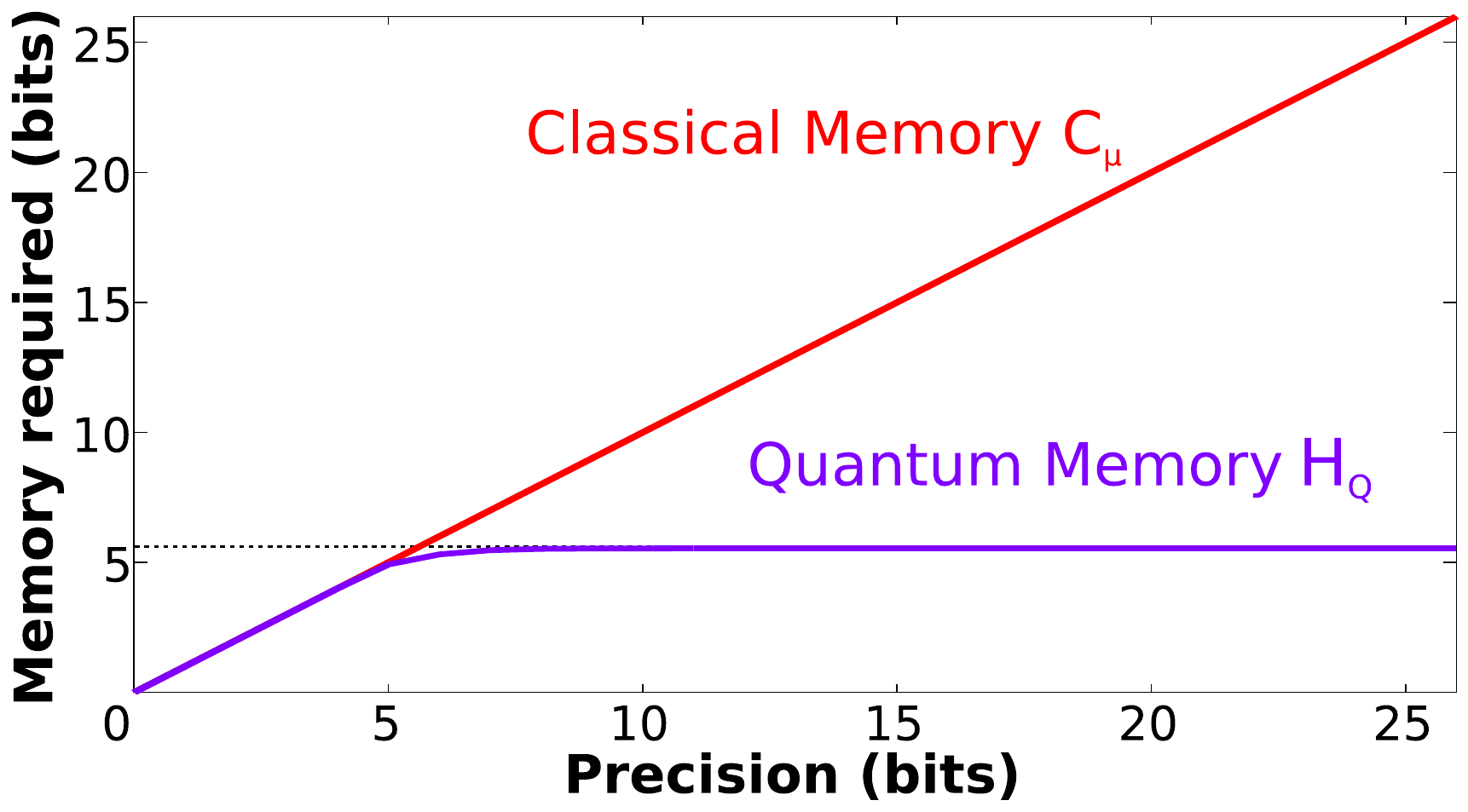}
\subcaption{\label{fig:GSlice} Gaussian noise function $\sigma = 0.01$, demonstrating unbounded difference in classical and quantum memory requirements.
The dotted line shows the analytic upper bound from eq.~\eqref{eq:GaussianCost}.
}
\end{subfigure}
\hspace{1em}
\begin{subfigure}[t]{0.4\textwidth}
\includegraphics[width=0.95\textwidth]{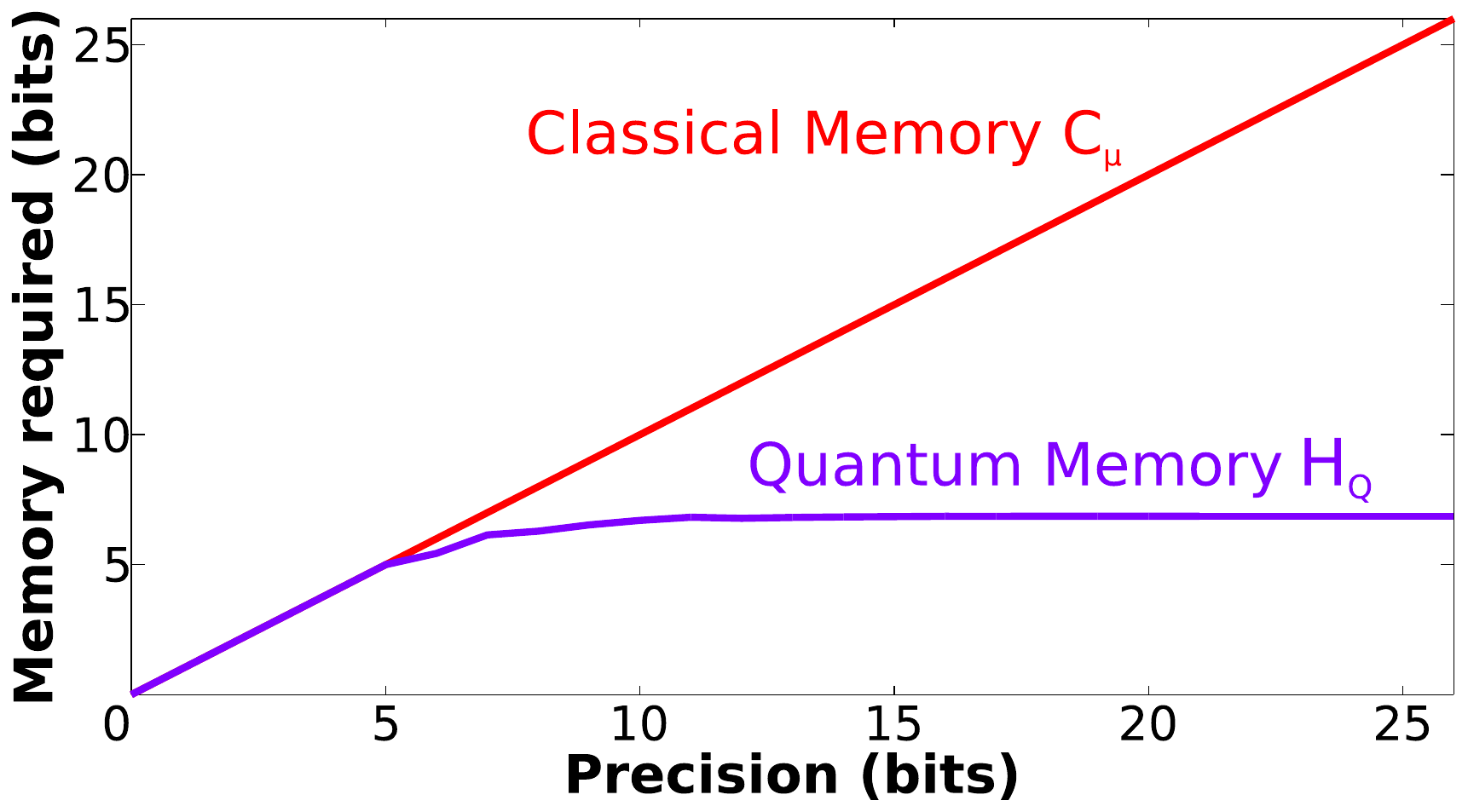}
\subcaption{\label{fig:THSlice}
Uniform noise function $\Delta = 0.01$, demonstrating unbounded difference in classical and quantum memory requirements.}
\end{subfigure}
\end{centering}\\
\caption{
\label{fig:Examples}
\caphead{Bounded quantum memory costs for unbounded precision.}
The memory required to simulate a cyclic random walk
 is plotted against the precision $N$ for the {\em Gaussian} and {\em top-hat} shift functions.
In both examples, the quantum simulator has an unbounded memory advantage -- the classical cost scales as $\log N$ whilst the quantum cost converges upon a constant value.
The more rapidly the shift function diffuses $X$, the lower the limiting quantum memory requirement.
}
\end{figure*}

Explicitly, we show that for certain cyclic processes, the quantum ensemble state's eigenvalues $\{\lambda_k\}_{k=0\ldots N-1}$ satisfy $\lim_{N\to\infty} \sum_{k=0}^{N-1} -\lambda_k \log \lambda_k = \Omega$ for some finite value $\Omega$.
Our result relies on first observing that the eigenvalues $\lambda_k$ can be directly related to transition probabilities $\{p_{jk}\}$ via the relation
\begin{equation}
\label{eq:EigenExact}
\lambda_k = \frac{1}{N} \mathcal{F}\!\left[\sqrt{p_{j0}}\right] \mathcal{F}\!\left[\sqrt{p_{(N-j)0}}\right],
\end{equation}
 where $\mathcal{F}$ denotes the {\em discrete Fourier transform}, $\Fo{x_j} = \sum_{j=0}^{N-1} x_j \exp\left(\frac{-2\pi i}{N} j k\right)$. (The proof relies on invoking the cyclic symmetry of the process -- and hence of the transition probabilities --  and is explicitly derived in the {\em Technical Appendix}).
The spread $p_{j0}$ (as a function of $j$) is an indicator of how quickly a particle diffuses in the random walk.
Thus, the Fourier-like relation between $p_{j0}$ and $\lambda_k$ indicates an inverse relationship between the amount of diffusion in the cyclic process and the spread of eigenvalues.
The greater the variance of $X$, the more quickly a particle diffuses, and the smaller the spread of $\lambda_k$ -- resulting in a reduced quantum memory requirement.
We now show that for some natural examples, this reduction is sufficiently large that $H_q$ remains bounded for all $n$ (as illustrated in \cref{fig:Examples}).

\inlinesubheading{Example 1: Gaussian noise.}
A cyclic process rotating at a constant rate subject to Gaussian noise has a shift function given by a Gaussian distribution $G_{\mu,\sigma}(x) = \frac{1}{\sigma\sqrt{2\pi}} \exp\left(-\frac{(x-\mu)^2}{2\sigma^2}\right)$ about mean $\mu$ with standard deviation $\sigma$.
Here, $\mu$ characterises the average velocity (in terms of the variable's mean displacement per time-step), and $\sigma$ the size of the fluctuations.
When $\mu=0$, this process corresponds to Gaussian diffusion.
For our analysis, we take $\sigma \ll 1$ and thus ignore fluctuations where the particle travels more than a complete loop around the ring in a single time-step (a value of $\sigma=0.1$ ensures that such events are less likely than one part in a million.)

\enlargethispage{\baselineskip}
As can be seen in \cref{fig:GPlot,fig:GSlice}, as the desired precision increases, the memory cost of simulating this process quickly converges onto a constant determined by the fluctuation strength $\sigma$; ultimately, \emph{infinite-precision \mbox{simulation} is possible using only a finite quantum memory}.
This behaviour may be understood analytically by seeing that for large $N$, the eigenvalues associated with the quantum simulator's internal memory are also given by samples from a Gaussian distribution:
$\lambda_k = G_{0,\frac{1}{4\pi\sigma}}(k)$ for $k = -\frac{N}{2}, \ldots \frac{N}{2}-1$,
where for convenience we have cyclicly offset the label of the eigenvalues' indices by $N$ (proof in {\em Technical Appendix}).
This demonstrates that increasing $\sigma$ tightens the spread of eigenvalues, and thus reduces the memory requirement for the quantum simulator.

In the {\em Technical Appendix}, we prove that as the precision $n = \log N$ increases, the sum $\lim_{N\to\infty} \sum_{k=-\frac{N}{2}}^{\frac{N}{2}-1} \lambda_k \log \lambda_k$ converges on a finite value, bounded (in bits) by
\begin{equation}
\label{eq:GaussianCost}
H_Q \leq \frac{1}{2\ln 2} - \left(1+4\sqrt{2\pi}\sigma\right) \log_2 2\sqrt{2\pi}\sigma
\end{equation}
Thus, for any fixed $0 <\sigma\ll 1$, the Gaussian random walk may be simulated to arbitrarily high precision using a quantum simulator of bounded entropy.
Moreover, this also implies an unbounded divergence between the classical and the {\em quantum statistical complexity}~\cite{SuenTGVG15,AghamohammadiMC16} $C_Q$, which is upper bounded by $H_Q$.

\inlinesubheading{Example 2: Uniform white noise.}
In the second example, we consider a particle that is perturbed by uniformly distributed noise.
At each time-step, the particle can move anywhere in the range of $\mu \pm \Delta$ from its current position with uniform probability, where $\Delta<\frac{1}{2}$.
Again, $\mu$ characterises the average velocity, and here $\Delta$ the size of the fluctuations.
The associated shift function is a {\em top-hat function}, that has a uniform value of $\frac{1}{2\Delta}$ in the range $x\in[\mu-\Delta, \mu+\Delta]$ and $0$ everywhere else.

The entropy of the quantum simulator, $H_q$ is plotted for various precision in \cref{fig:THPlot,fig:THSlice}.
We see that for any fixed $\Delta>0$, the quantum memory required by our simulator converges to a bounded value.
As in the Gaussian scenario, the quantum simulator can replicate a classical simulation to any given precision using with finite entropy.
In the {\em Technical Appendix}, we prove this analytically.
We show that as $N \rightarrow \infty$, the entropy remains finite, and is bounded above by $H_{Q} \leq \frac{1.894}{\sqrt{\Delta}} + 3.067$.
In particular, for large $N$, the eigenvalues of the relevant ensemble state obey $\lambda_k = 2\Delta \sinc^2\!\left(2 k \Delta \right)$ for $k = -\frac{N}{2}, \ldots \frac{N}{2}-1$, where $\sinc(x)$ is the {\em normalized sinc function}, $\sinc(x) := \frac{1}{\pi x} \sin(\pi x)$. Larger values $\Delta$ will result in a smaller spread of eigenvalues, and result is smaller $H_q$.
For any given $\Delta>0$ the entropy is finite in the limit $N\to\infty$. 
This establishes a second natural example where the quantum simulator can demonstrate an unbounded memory advantage over its best possible classical counterpart.

\inlineheading{The origin of quantum advantage.}
The source of classical inefficiency can be understood by considering dynamics on causal states.
Consider two instances of $\mathcal{P}_n$, 
 one where $Y^0 = y_j$, and the other where $Y^0 = y_{j+1}$.
As their conditional future statistics differ [that is, $\cPrline{\future{X}}{Y^0 = y_j}\neq \cPrline{\future{X}}{Y^0 = y_{j+1}}$], a classical simulator must be configured differently for each instance (corresponding to being initialized in one of two different causal states, $s_j$ or $s_{j+1}$).
Nevertheless, there is finite probability that at the next time-step, both instances of the process emit the same output (up to precision $n$).
Should this happen, we would not be able to use the current state of the machine to determine the causal state it was in at the previous time.
That is, there is some probability that the distinction between $s_j$ and $s_{j+1}$ will never be reflected in the future statistics of the process -- a phenomenon known as crypticity~\cite{MahoneyEC09,MahoneyAC16}.
As $n$ increases, this occurs with greater likelihood (tending to unit probability as $n \rightarrow \infty$), and thus proportionally more information is wasted.
Ultimately, in the limit of high precision, a vanishingly small proportion of the information stored in the classical memory is pertinent to the statistical behaviour of the process's future.

Quantum simulators compensate for this waste by mapping these causal states to non-orthogonal quantum states.
The quantum state (\cref{eq:QStates}) associated with neighbouring causal states ($\ket{S_j}$ and $\ket{S_{j+1}}$) also become increasingly similar with increasing $n$ -- resulting in progressively greater savings.
Consider the Gaussian scenerio, where $H_q$ is bounded by equation~\eqref{eq:GaussianCost}.
 For small $\sigma$, the memory cost scales as $-\log_2 \sigma$, such that halving the variance of fluctuations at each time-step adds one bit to the memory cost of the quantum simulator.
The standard deviation of the shift function has set an effective length scale over which the system must be simulated classically.
The statistical behaviour of future outputs from two systems that are initially prepared in points separated by more than one standard deviation are typically distinguishable, and so these points must be stored as nearly-orthogonal quantum states at some memory cost.
On the other hand, when two points are initially closer than the standard deviation scale, the probability that they could be distinguished by their future behaviour diminishes, and they may be represented by increasingly overlapping quantum states.
In this regime, a fixed finite memory can accommodate any desired precision.

We gain further insight into the origins of quantum advantage by considering the cases where it does not appear: $\sigma=0$ and $\Delta=0$.
In both these cases, the shift function is a Dirac delta distribution.
As such, no matter how high the precision, by observing the future outputs, it will {\em always} be possible to distinguish whether the system came from some site $s_j$ or its neighbour $s_{j+1}$; the dynamics of the system are wholly reversible.
If $s_j$ always transitions to $s_k$ and $s_{j+1}$ always to $s_{k+1}$, being able to distinguish between these two sites is crucial to produce the correct statistical behaviour, even as the precision increases.
As such, the quantum simulator cannot tolerate overlap between the states $\ket{s_j}$ and $\ket{s_{j+1}}$, and must store them orthogonally (allowing them to be distinguished).
In this scenario, the quantum simulator cannot demonstrate any advantage in memory cost over its classical analogue.

\inlineheading{Discussion and outlook.}
In this article, we presented a task in which quantum mechanics has an unbounded memory advantage over the most memory-efficient classical alternative:
 the simulation of a classical cyclic stochastic process.
We found that the classical simulator has a memory requirement that scales
linearly with the precision required, while the quantum simulator's requirement may be bounded by a finite value, even at arbitrarily-high fixed precision.
This establishes a rare scenario where the scaling advantage of quantum processing can be provably established.

This finding leads to a number of natural open questions -- the first being of generality.
Certainly, the examples presented are sufficiently simple that such divergences are unlikely to be merely a mathematical oddity.
The unbounded quantum advantage relies on $\{\mathcal{P}_n\}$ having two properties: (a) the number of causal states grows with $n$, and (b) the conditional future statistics $\cPrline{\future{X}}{S=s_i}$ between different causal states converges sufficient quickly with $n$.
If these conditions can be formalized, we may be able to establish similar divergences in much more general scenarios,
 such as the simulation of non-Markovian or non-cyclic processes.
Beyond von Neumann entropy, it would be interesting if similar scaling can be found for other metrics of memory cost, such as the dimension -- namely, whether there is an encoding that allows for simulation to arbitrary precision using a Hilbert space of bounded dimension.
Meanwhile the inefficiency of classical simulators have show to directly results in unavoidable increased heat dissipation~\cite{WiesnerGRV12,StillSBC12,GarnerTVG15}.
This hints that quantum processing may allow significant energetic savings for stochastic simulation, 
 especially for systems that become increasingly difficult to simulate as they scale in size.

On a foundational level, the statistical complexity is often regarded as a fundamental measure of a process's intrinsic structure 
 --  the rationale being that it quantifies the minimal amount of information about a process's history that must be recorded to allow for predictions about that process's future behaviour.
The measure has been applied to understand structure within diverse complex settings: from the dynamics of neurons~\cite{HaslingerKS10} and the stock market~\cite{ParkWLYJM07}, to quantifying self-organization~\cite{ShaliziSH04}, among other examples~\cite{MarquesdaSilva1997,ClarkeFW03,LiYK08,LuB12}.
The discovery of more efficient quantum models has led to the idea that the complexity of a system depends on what sort of information we use to observe it~\cite{SuenTGVG15,AghamohammadiMC16}.
In this context, our results establish a family of processes that can look ever more complex classically, but remain simple quantum-mechanically.
It would fascinating to see if divergences between quantum and classical complexities can be found in existing studies, such as the examples above.
Could it be that these systems appear complex classically -- but look much simpler when viewed through the lens of quantum theory?

\section*{ACKNOWLEDGEMENTS}
We thank James Crutchfield, Thomas Elliott, David~Garner, Peter Grassberger, \mbox{Jan-\AA{}ke} Larsson, and Chengran~Yang for helpful comments and discussions.
We gratefully acknowledge funding from
 the John Templeton Foundation Grant 53914 {\em ``Occam's Quantum Mechanical Razor: Can Quantum theory admit the Simplest Understanding of Reality?''};
 the Foundational Questions Institute;
 the Ministry of Education in Singapore, the Academic Research Fund Tier 3 MOE2012-T3-1-009; and the
 the National Research Foundation of Singapore (Award Nos.\ NRF--NRFF2016--02 and NRF--CRP14-2014-02).

\section*{TECHNICAL APPENDIX}

\inlineheading{Classical costs from computational mechanics.}
We here present some minimal details from the mathematical framework of computational mechanics~\cite{CrutchfieldY89,ShaliziC01,CrutchfieldEM09,Crutchfield11} to substantiate the claim that the classical simulator's minimal memory cost is equal to the precision $\log N$.

In computational mechanics,
 the evolution of a dynamical property (over domain $\mathcal{Y}$) is characterised by a discrete-time stochastic process $\mathcal{P}$,
 written as bi-infinite sequence of random variables $\{Y^t\}_{t\in\mathbb{Z}}$,
 where each random variable $Y^t$ governs the value $y^t\in\mathcal{Y}$ of the dynamical property at time $t$.
The statistical behaviour of a process
 may be represented in a {\em causal} manner
 by writing it as the conditional probability distribution $\cPrline{\future{Y}_t}{\past{Y}_t}$,
 where $\future{Y}^t = Y^{t+1}Y^{t+2}\ldots$ is the infinite string of random variables occuring after time $t$,
 and $\past{Y}^t = \ldots Y^{t-1} Y^t$ is the infinite string of random variables occuring before (and including) time $t$.
For {\em stationary} processes (such as the time-independent cyclic random walks described in this article), this distribution has no explicit time dependence, so we omit the superscript $t$.

A faithful {\em simulator} of process $\mathcal{P}$ is
 a machine (or program) that, having been initialized in accordance with the observation of past $\past{y}^t$,
 then generates a series of outputs $\future{y}_t$ according to the distribution $\cPrline{\future{Y}^t = \future{y}^t}{\past{Y}^t=\past{y}^t}$.
Since storing an infinite string $\past{y}^t$ may require an unbounded amount of memory,
 one instead configures the internal state of the simulator $s$ (over configuration space $\mathcal{S}$) according to some function $s = f\!\left(\past{y}\right)$, satisfying
  $\cPrline{\future{Y}^t = \future{y}^t}{S = s} = \cPrline{\future{Y}^t = \future{y}^t}{\past{Y}^t=\past{y}^t}$,
 where $S=f\small(\past{Y}\small)$ is the random variable describing the internal state of the simulator (formed by applying the function $f$ on each variate of $\past{Y}$).
Moreover, once initiated into state $s^t$, when the simulator outputs $y^t$ in the subsequent time-step, its internal state must then transition to the state $s^{t+1} = f(\past{y} y^t)$ (where $\past{y} y^t$ indicates the concatenation of $y^t$ to the end of string $\past{y}$).

The memory cost of such a simulator is given by the information entropy of $S$, $\Ent{S} = -\sum_{s_i\in\mathcal{S}} \Pr{S\!=\!s_i} \log \Pr{S\!=\!s_i}$.
The function $f$ that minimizes this classically corresponds to identifying the {\em causal state} of a particular past~\cite{CrutchfieldY89,ShaliziC01},
defined by the equivalence relationship: $\past{y} \sim_\epsilon \past{y}'$ for pasts $\past{y}$ and $\past{y}'$ if and only if $\cPrline{\future{Y} = \future{y}}{\past{X}_t = \past{y}} = \cPrline{\future{Y} =\future{y}}{\past{X}_t = \past{y}'}$ for all possible future values $\future{y}\in\future{Y}$.
The causal states are unique for any given process, and so their entropy $\Ent{S}$ is a property of the process itself known as its {\em statistical complexity} $C_\mu$, capturing the intuition that a more complex process requires more memory to simulate.

For Markovian processes, such as discussed in this article,
 the number of causal states required is equal to the number of unique rows in the stochastic matrix describing the evolution.
When these rows are generated by the discretization of a continuous process into $N$ divisions -- such as when they are derived from the cyclic walk's shift function $P(X)$ --
 the number of states will be equal to $N$, except for very specific (e.g.\ pathologically fractal) choices of $P(X)$ and $N$.
Since by symmetry the probability of the simulator being in any particular state is equal,
 the classical memory cost of a simulator hence scales with the number of sites as $\log N$, or linearly with the precision $n = \log_2 N$.

\inlineheading{Details of the quantum circuit in \cref{fig:Circuit}.}
Let us consider \cref{fig:Circuit} in more depth (see also \cite{GuWRV12}).
The circuit consists of one persistent internal memory state, and an ``output tape''---a line of quantum states, which are fed into the system one at a time.
Suppose each state on the output tape is initialized into some arbitrary state $\ket{\phi}$.
For any two quantum states $\ket{x}$ and $\ket{y}$ in the same Hilbert space,
 it is always possible to construct a unitary transformation $V$ such that $V \ket{x} = \ket{y}$.
This will be of the form $\ketbra{y}{x} + \sum_i \ketbra{y_i'}{x_i'}$ where $\ket{x_i'}$ are states orthogonal to each other and to $\ket{x}$, and $\ket{y_i'}$ are states orthogonal to each other and to $\ket{y}$.
Thus, in the joint Hilbert space $\mathcal{H}^N\otimes\mathcal{H}^N$ of two quantum systems of dimension $N$,
 it is possible to build a ``controlled'' unitary operation $U$ containing the elements $\ketbra{j}{j} \otimes \ketbra{\psi_j}{\phi}$ for every $\ket{\psi_j}$ in an arbitrary (generally non-orthogonal) set of states $\{\psi_j\}_{j=0\ldots \left(N-1\right)}$.
[Note: the orthogonality of $\{\ket{j}\}$ allows us to pairwise use the above construction for each $\ket{\psi_j}$.]

For a Markovian process discretized such that the stochastic matrix with elements $p_{jk}$ describes its evolution,
 the above prescription supplies the unitary operation required for our quantum simulator
 when we set each $\ket{\psi_j} = \ket{S_j} = \sum_{k=0}^{N-1} \sqrt{p_{kj}}\ket{k}$, as per \cref{eq:QStates} (states $\{\ket{k}\}$ and $\{\ket{j}\}$ are in the same basis).

We may now evaluate the action of a single time-step (grey dashed box within \cref{fig:Circuit}).
Here, the joint Hilbert space corresponds to that of the internal memory together with the output tape.
In the figure, we explicitly wrote the initial state of the output tape as $\ket{\phi} = \ket{S_0}$, but this is arbitrary;
 any $\ket{\phi}$ could be made into $\ket{S_0}$ by acting on it first with a unitary gate containing $\ketbra{S_0}{\phi}$.
At the start of a time step, the internal memory is in state $\ket{S^t} = \ket{S_j} =\sum_{k=0}^{N-1} \sqrt{p_{kj}}\ket{k}$.
Hence, the joint state of the memory and output tape is initially $\ket{S_j}\otimes\ket{\phi}$.
After the controlled unitary is applied, the memory and tape will be in the entangled state $\sum_k \sqrt{p_{kj}} \ket{k}\otimes\ket{S_k}$.
Applying a coherent swap operation (i.e.\ exchanging the labels of the Hilbert spaces) will take this joint state to $\sum_k \sqrt{p_kj} \ket{S_k}\otimes\ket{k}$ -- the state of the system at the end of the grey box.

The tape system is then ejected from the simulator.
If one were to measure this state in the $\{\ket{k}\}$ basis, one projects onto state $\ket{k}$ with probability $p_{kj}$,
 and hence the output statistics of this measurement match that of the process being simulated.
Moreover, after measuring, due to the entanglement, we know that when $\ket{k}$ is measured, the internal memory must be in state $\ket{S_k}$,
 which is exactly the quantum state that would have been prepared if we had mapping the output statistics onto a classical causal state and then prepared $\ket{S_k}$ directly.
Hence, the quantum circuit in \cref{fig:Circuit} can function as a discretized simulator for a Markovian process.

However, it is very important to note that there is no need whatsoever to measure the output tape $\ket{k}$ for the quantum simulator to continue functioning.
If it suits one's purpose to store the output states in quantum memory (e.g.\ to perform further quantum information processing on the output data), then the quantum simulator still functions correctly.
In this mode of operation, the measurements can be omitted from \cref{fig:Circuit}, and after $M$ steps, the simulator would have produced the entangled state
\begin{align}
\ket{\Phi} = &
\sum_{i_1} \ldots \sum_{i_M} \sqrt{P(Y^t\!=\!y_{i_1}, \ldots Y^{t+M}\!=\!y_{i_M} | S^t)} \nonumber \\
& \quad \ket{S^{t+M}(S^t, y_{i_1}, \ldots y_{i_M})} \otimes \ket{y_{i_1}} \otimes \ldots \ket{y_{y+M}}
\end{align}
where $\ket{S^{t+M}(S^t, y_{i_1}, \ldots y_{i_M})}$ is the quantum state that would have been prepared if the system was originally in causal state $S^t$ then outputted string $y_{i_1}\ldots y_{i_M}$, and a new causal state directly set according to this output sequence.
Measuring the string of output tape subsystems thus still ensures that the internal memory state collapses into the correct causal state $\ket{S^{t+M}}$, conditional on the string observed.

In the first mode of operation (as drawn in \cref{fig:Circuit}), only one ancillary quantum system is required, as it can be reset and re-used between timesteps (the output tape carries away classical information only).
In the second mode, the quantum output explicitly fulfils the role of the ancillary system, and a fresh ancillary system (provided by the ``blank'' output tape set to some fixed choice of pure quantum state) is inserted at each time step.
In both modes, the ancillary system does not need to persist between time steps in order for the simulator to continue producing statistically correct outputs.
As such, in both cases, it is the von Neumann entropy $-\tr \rho\log\rho$ of the first subsystem, which remains within the simulator at all times, that we consider to be the internal memory cost.

\inlineheading{Derivation of discrete eigenspectrum.}
The quantum machine state corresponding to the system being in classical state $\alpha$ is given as $\ket{S_\alpha} = \sum_\beta \sqrt{ p_{\beta\alpha} } \ket{\beta}$.
Assuming $\{p_{\beta\alpha}\}_{\beta\alpha}$ is simply connected,
 the quantum machine will reach a stationary state $\rho = \frac{1}{N} \sum_\alpha \ketbra{S_\alpha}{S_\alpha}$.
Rather than directly calculating the entropy of $\rho$, we can instead evaluate the entropy of the associated {\em Gram matrix} $g$, whose elements $g_{\alpha\beta}$ are given by the overlaps $\frac{1}{N} \braket{S_\alpha}{S_\beta}$.\footnote{
This works by constructing a fictitious purification of $\rho$, given $\ket{\Psi} = \sum_i \frac{1}{\sqrt{N}} \ket{S_i}\otimes\ket{i}$ (where $\{\ket{i}\}_i$ is an orthonormal basis) such that $\tr_B \ketbra{\Psi}{\Psi} = \rho$ and $\tr_A \ketbra{\Psi}{\Psi} = g$. Since the von~Neumann entropy of pure state $\ket{\Psi}$ is $0$, it follows from triangle inequalities that $\Ent{\rho} = \Ent{g}$.
}
The circular symmetry of the cyclic random walk ensures that the discretized transition probabilities satisfy $p_{\alpha\beta} = p_{(\alpha+k)(\beta+k)}$ (that is, the transition probabilities depend only on differences between indices). 
It hence follows that the Gram matrix associated with $\rho$ is {\em circulant}~\cite{Gray05}.
Since all rows can be derived by cyclic permutation of the top row, we shall drop one index and write the top row as $g_\alpha = g_{0\alpha}$.
The eigenvalues of the Gram matrix are given by $\lambda_k = \sum_\alpha g_{\alpha} \exp{\left(-\frac{2\pi i}{N} \alpha k\right)}$ for $k=0, \ldots, N-1$,
 which can immediately be recognized as the discrete Fourier transform (DFT) of $\{g_{\alpha}\}_\alpha$, which we denote as $\mathcal{F}(g_{0\alpha})$.

Moreover, the inner product $\braket{S_0}{S_j} = \sum_\alpha \sqrt{p_{\alpha0} p_{\alpha j}}$, has the form of a convolution $\sqrt{p}\ast\!\sqrt{q}$, where we have rewritten $p_{\alpha j}$ as $q_{0(\alpha-j)}$ such that $q$ is the $N$-periodic extension of the reflection of $p$; $q_{0j} = p_{(N-j)0}$ and $q_{0j} = q_{0(j+N)}$.
We may then apply the circular convolution theorem to
 find the eigenvalues of $g$, and therefore of~$\rho$:
\begin{equation}
\label{eq:EigenExactApp}
\lambda_k = \frac{1}{N} \mathcal{F}\!\left[\sqrt{p_{j0}}\right] \mathcal{F}\!\left[\sqrt{p_{(N-j)0}}\right].
\end{equation}
These eigenvalues can hence be found efficiently by numerical algorithms, such as the fast-Fourier transform.

\inlinesubheading{Example: Dirac-delta shift function.}
Let the shift function be $P(x) = \delta(x - x_0)$ for some $x_0 \in [0,1)$.
It can be seen that all $p_{j0} = 0$ except for the one at index $j'$ that incorporates the delta peak where $p_{j'0} = 1$.
Hence, $\Fo{p_{j0}} = \exp\left(-2\pi i \frac{j'}{N} k\right)$ and $\Fo{p_{(N-j)0}} = \exp(-2\pi i \frac{\left(N-j'\right)}{N} k)$, and so $\lambda_k = \frac{1}{N}$ for all~$k$.
Thus, the von Neumann entropy of the simulator's memory is $\log N$.

\inlinesubheading{Example: Uniform shift function.}
Consider the uniform shift function $P(x)=1$ for $x\in[0,1)$.
Here, $p_{j0} = \frac{1}{N}$, and so $\mathcal{F}\!\left[\sqrt{p_{j0}}\right] = \sqrt{N}$ for $k=0$ and $0$ for all other $k$.
As such, we find that the eigenvalue $\lambda_0 = 1$, and all other eigenvalues $\lambda_1 = \ldots \lambda_{N-1} = 0$, and hence the entropy of the Gram matrix is zero, for all values of~$N$.

\inlineheading{Sampling Fourier transforms.}
It will be useful to show an auxiliary relationship between discrete and continuous Fourier transforms.
Let $g(x)$ be a function over the range $x\in[0,1]$ that is sampled at $N$ equally spaced points with values given by $g_n = g(\frac{n}{N})$ for $n=0\ldots N-1$.
We can construct a function $g_{\rm comb}(x) = \sum_{n=0}^{N-1} \delta\!\left(x - \frac{n}{N}\right) g(x)$, whose Fourier transform is
\begin{align}
\Fo{g_{\rm comb}(x)} & = \int_{-\infty}^{\infty} dx \, \sum_{n=0}^{N-1} \delta\!\left(x - \frac{n}{N}\right) g(x) \exp{\left(-2\pi i k x\right)} \nonumber \\
& = \sum_{n=0}^{N-1} g\left(\frac{n}{N}\right) \exp{\left(-2\pi i \frac{n}{N} k\right)},
\end{align}
which when evaluated at integer $k$ is exactly the DFT of the samples $\{g_n\}$, which we write as $\{\lambda_k\}$.

If $g$ is periodic, it is always possible to offset the position of the sample window of $g$ by some integer $c$ without changing the values of $g$'s DFT.
For the functions we consider in this article, it is more convenient to start at $-\frac{N}{2}$, since typically $g_{-\frac{N}{2}}, g_{\frac{N}{2}} \to 0$ and $g_0 = 1$.
Moreover, once the sample window has been set, the values of $g(x)$ outside this window can not affect $\lambda_k$, since they do not feature in the sum.
Thus, instead of considering sampling $g(x)$ across a finite window, we can consider an infinite delta train sampled at the same intervals, but across a function $g_{\rm once}(x)$ where $g_{\rm once}(x) = g(x)$ inside the range of the sample window (i.e. $[-\frac{1}{2}, \frac{1}{2})$ for the window used in this article) and $g_{\rm once}(x)=0$ outside this range.
Here
\begin{align}
\lambda_k = \Fo{g_{\rm comb}(x)} & = \Fo{
g(x) \sum_{n=-\frac{N}{2}}^{\frac{N}{2}} \delta(x - \frac{n}{N})
} \nonumber \\
& = \Fo{
g_{\rm once}(x) \sum_{n=-\infty}^{\infty} \delta(x - \frac{n}{N})
} \nonumber \\
\label{eq:ContTrain}
& = \Fo{g_{\rm once}(x)} \conv \sum_{m=-\infty}^{\infty} \delta(k - mN),
\end{align}
where we have used the convolution theorem in the final step.
The periodic sampling of $g(x)$ causes the Fourier transform to be periodic with period $N$ (a phenomenon known as aliasing), such that $\lambda_k = \lambda_{k+N}$; the convolution with a delta train effectively makes $\lambda_k$ a {\em periodic sum} of $\Fo{g_{\rm once}(x)}$.
This periodicity allows us the freedom to choose a convenient range of $k$.
In this article, we will typically use $-\frac{N}{2}$ to $\frac{N}{2}-1$.
If $\Fo{g_{\rm once}(x)}\approx 0$ outside the chosen range, then we can approximate
\begin{equation}
\lambda_k \approx \left[\Fo{g_{\rm once}(x)}\right](k).
\end{equation}

\inlineheading{Asymptotic limit of eigenvalues.}
For large $N$, we can derive an expression for $\lambda_k$ in terms of the probability density function $P(x)$.
We substitute $p_{\alpha 0}$ with $\frac{1}{N} P(\frac{\alpha}{N})$, which for Riemann-integrable $P(x)$ is an arbitrarily good approximation in the limit of $N\to\infty$.
Similarly, we may substitute $p_{\alpha j}$  with $\frac{1}{N} P^\circ(-\frac{j-\alpha}{N})$, where $P^\circ(x)$ denotes the $1$-periodic extension\footnote{Equivalent to wrapping $x$ to $[0,1)$ before evaluating $P(x)$.} of $P(x)$.
Taking the limit of the Riemann sum for a product of two functions, we then see
\begin{align}
\lim_{N\to\infty} \braket{S_0}{S_j} & = \lim_{N\to\infty} \sum_{\alpha=0}^{N-1} \sqrt{p_{\alpha 0} p_{\alpha j}} \nonumber \\
& = \lim_{N\to\infty} \sum_{\alpha=0}^{N-1} \frac{1}{N} \sqrt{P(\frac{\alpha}{N}) P^\circ(-\frac{j-\alpha}{N})} \nonumber \\
& = \int_0^1 dx \sqrt{ P(x) P^\circ (y-x)},
\end{align}
where $y = \frac{j}{N}$.
Moreover, since $P$ only has support in $[0,1)$, we can rewrite the integral limits from $-\infty$ to $\infty$, and conclude that $\lim_{N\to\infty} \braket{S_0}{S_j} = [\sqrt{P(x)}\conv\sqrt{P^\circ(-x)}](y)$ sampled at $y= 0, \frac{1}{N}, \ldots \frac{N-1}{N}$.
Thus by treating $g_j$ as samples from a function $g(y=\frac{j}{N})$ at discrete intervals of $\frac{1}{N}$, we find that $g_j \approx \frac{1}{N} g(y=\frac{j}{N})$ for large $N$, and hence
\begin{equation}
\label{eq:AsymGram}
g(y=\frac{j}{N}) = \left[\sqrt{P(x)}\conv\sqrt{P^\circ(-x)}\right]\left( y \right).
\end{equation}

As shown in eq.~\eqref{eq:ContTrain}, the eigenvalues $\{\lambda_k\}$ are given by
 $\lambda_k = \left[\Fo{g_{\rm once}(x)}\conv \sum_{m=-\infty}^{\infty} \delta(k - mN)\right]$
 evaluated at integers $k=0, 1, \ldots N-1$, where $g_{\rm once}(y) = g(y)$ over an (arbitrary) single period of $g(y)$ and takes the value zero elsewhere.
Due to the periodic summation, it can be seen also that $\lambda_k = \lambda_{k+N}$, and so we are also free to choose the most convenient range for $k$, which will typically be from $-\frac{N}{2}$ to $\frac{N}{2}-1$.
If $[\Fo{g_{\rm once}}](k) \approx 0$ when $|k|>\frac{N}{2}$, then the approximation
\begin{equation}
\label{eq:AsymEigen}
\lambda_k \approx [\Fo{g_{\rm once}}](k) \quad \mathrm{for~} k=-\frac{N}{2}, \ldots \frac{N}{2}-1
\end{equation} is reasonable.
This assumption amounts taking enough samples of $g(x)$ to admit a faithful reconstruction of $g(x)$ under the Nyquist--Shannon theorem~\cite{Shannon49}. 
This holds true for the examples we shall now consider, where we will ultimately take large values of $N$.

\inlineheading{Example 1: Gaussian noise.}
Suppose the shift function of the particle is given by a Gaussian distribution $G_{\mu,\sigma}(x) = \frac{1}{\sigma\sqrt{2\pi}} \exp\left(-\frac{(x-\mu)^2}{2\sigma^2}\right)$ about $\mu$ with standard deviation $\sigma \ll 1$ such that we can ignore the probability of the particle looping around the ring.

\inlinesubheading{Derivation of eigenvalues.}
We can express $\sqrt{G_{\mu,\sigma}(x)}$ as a Gaussian:
\begin{align}
\sqrt{G_{\mu,\sigma}(x)} &= \sigma^{-\frac{1}{2}}\left(2\pi\right)^{-\frac{1}{4}} \exp\left(-\frac{(x-\mu)^2}{4\sigma^2}\right) \nonumber \\
& = \sigma^{\frac{1}{2}}\left(2\pi\right)^{\frac{1}{4}} \sqrt{2} (\sqrt{2}\sigma)^{-1}\left(2\pi\right)^{-\frac{1}{2}} \exp\left(-\frac{(x-\mu)^2}{2(\sqrt{2}\sigma)^2}\right) \nonumber \\
& = \sigma^{\frac{1}{2}} 2^{\frac{1}{2}}\left(2\pi\right)^{\frac{1}{4}} G_{\mu,\sqrt{2}\sigma}(x).
\end{align}

It can be easily verified that $g_{\mu,\sigma}(-x) = g_{-\mu,\sigma}(x)$.

We also note that $\Fo{g_{\mu,\sigma}(x)}$ is also Gaussian:
\begin{align}
\Fo{g_{\mu,\sigma}(x)} & = \exp\left(2\pi i \mu k\right) \exp\left(-2(\pi\sigma)^2 k^2\right) \nonumber \\
& \hspace{-3em} = (2\pi)^{\frac{1}{2}} \frac{1}{2 \pi \sigma} \exp\left(-2\pi i \mu k\right) \nonumber \\
& \hspace{3em} \cdot (2\pi)^{-\frac{1}{2}}  (\frac{1}{2 \pi \sigma})^{-1}
\exp\left(-\dfrac{k^2}{2\left(\frac{1}{2\pi\sigma}\right)^2}\right) \nonumber \\
& =  (2\pi)^{-\frac{1}{2}} \sigma^{-1} \exp\left(-2\pi i \mu k\right) g_{0,\frac{1}{2\pi\sigma}}(k)
\end{align}

Likewise, we can express $[G_{\mu,\sigma}(x)]^2$ as a Gaussian:
\begin{align}
[G_{\mu,\sigma}(x)]^2 &= \sigma^{-2}\left(2\pi\right)^{-1} \exp\left(-\frac{(x-\mu)^2}{\sigma^2}\right) \nonumber \\
& \hspace{-3em} = \sigma^{-1}\left(2\pi\right)^{-\frac{1}{2}} 2^{-\frac{1}{2}} (\frac{\sigma}{\sqrt{2}})^{-1} \left(2\pi\right)^{-\frac{1}{2}} \exp\left(-\frac{(x-\mu)^2}{2(\frac{\sigma}{\sqrt{2}})^2}\right) \nonumber \\
& \hspace{-3em} = \sigma^{-1}\left(2\pi\right)^{-\frac{1}{2}} 2^{-\frac{1}{2}} G_{\mu,\frac{\sigma}{\sqrt{2}}}(x).
\end{align}

Taken together (making sure to substitute in the correctly modified values of $\mu$ and $\sigma$), this allows us to provide an analytic solution for eq.~\eqref{eq:AsymEigen} for Gaussian shift functions:
\begin{align}
\lambda_k & =
 \Fo{\sqrt{G_{\mu,\sigma}(x)}}\Fo{\sqrt{G_{\mu,\sigma}(-x)}} \nonumber \\
& = \Fo{\sqrt{g_{\mu,\sigma}(x)}}\Fo{\sqrt{G_{-\mu,\sigma}(x)}} \nonumber \\
& = 2\sigma (2\pi)^\frac{1}{2} \Fo{G_{\mu,\sqrt{2}\sigma}(x)}\Fo{G_{-\mu,\sqrt{2}\sigma}(x)} \nonumber \\
& = 2\sigma (2\pi)^\frac{1}{2} (2\pi)^{-1} (\sqrt{2}\sigma)^{-2} \exp\left(-2\pi i \mu k\right)  \exp\left(2\pi i \mu k\right) \nonumber \\
& \hspace{3em} \cdot [G_{0,\frac{1}{2\sqrt{2}\pi\sigma}}(k)]^2 \nonumber \\
& = (2\pi)^{-\frac{1}{2}} \sigma^{-1} (\frac{1}{2\sqrt{2}\pi\sigma})^{-1} (2\pi)^{-\frac{1}{2}} 2^{-\frac{1}{2}} G_{0, \frac{1}{4\pi\sigma}}(k) \nonumber \\
&\label{eq:GaussTransApp} =  G_{0, \frac{1}{4\pi\sigma}}(k).
\end{align}

Hence, we see that choosing Gaussian transfer function with standard deviation $\sigma \ll 1$ corresponds to a spectrum of eigenvalues with standard deviation $\frac{1}{4\pi\sigma}$.

\inlinesubheading{Upper bound on quantum memory cost.}
We now demonstrate that the entropy of such a system, given $H_Q = -\sum_k \lambda_k \log_2 \lambda_k$, is finite by bounding it from above.
For convenience, we write $\lambda(k) := G_{0, \frac{1}{4\pi\sigma}}(k) = A \exp\left(-B k^2 \right)$ where
$A = 2\sqrt{2\pi}\sigma$ and $B=8 \pi^2 \sigma^2$,
 and will perform the calculation in units of {\em nats}.
Thus, consider $c(k) = -\lambda(k) \ln \lambda(k)$, explicitly
\begin{equation}
c(k) = A \exp\left(-Bk^2\right)\left(Bk^2-\ln A\right).
\end{equation}
By setting $\frac{dc}{dk} = 2 A B k \exp\left(-Bk^2\right)\left(- Bk^2 + \ln A + 1\right) = 0$,
 we find that $c(k)$ has stationary points at $k=0$, $\pm \infty$ and when
\begin{align}
\label{eq:MacCauch}
k & = \; \pm \sqrt{\dfrac{\ln A + 1}{B}}
 = \; \pm \sqrt{\dfrac{\ln\left(2\sqrt{2\pi}\sigma\right) +  1}{8\pi^2\sigma^2}}.
\end{align}
When $\sigma < \frac{1}{2 e \sqrt{2\pi}} \approx 0.073$, these last two solutions disappear,
 and since we are in the regime of $\sigma \ll 1$, this condition is satisfied.
Hence, for small $\sigma$, $c(k)$ monotonically decreases from its maximum value at $k=0$ for both positive and negative $k$.
This allows us to apply the Maclaurin--Cauchy integral bound (see e.g.~\cite{Knopp90}),
\begin{equation}
\int_m^\infty c(k) dk \leq \sum_{k=m}^\infty c(k) \leq c(m) + \int_m^\infty c(k) dk,
\end{equation}
which holds for any monotonically decreasing region $[m,\infty)$ of a function $c(k)$ (here, $m=0$).

Using known results for definite Gaussian integrals,
\begin{equation}
\int_0^\infty e^{-B x^2}dx = \frac{1}{2}\sqrt{\dfrac{\pi}{B}}
\quad\mathrm{and}\;
\int_0^\infty x^2 e^{-Bx^2}dx = \frac{1}{4}\sqrt{\dfrac{\pi}{B^3}},
\end{equation}
 we evaluate
\begin{align}
\int_0^\infty c(k) dk & = A B \frac{1}{4} \sqrt{\frac{\pi}{B^3}} - A \ln A \sqrt{\frac{\pi}{B}} \nonumber \\
& = \frac{A}{2}\sqrt{\frac{\pi}{B}}\left(\frac{1}{2} - \ln A\right) \nonumber \\
& =  \frac{1}{2}\left(\frac{1}{2} - \ln A\right).
\end{align}
Since $c(0) = -A \ln A$, we find from equation~\eqref{eq:MacCauch} that
\begin{align}
\sum_{k=m}^\infty c(k) &  \leq \left(\frac{1}{2} - \ln A\right) -A \ln A, \nonumber \\
 & \leq \frac{1}{4} - \left(\frac{1}{2}+A\right)\ln A .
\end{align}
To obtain a bound on $H_Q$, we double the above since $c(k)$ is even, and multiply by $\frac{1}{\ln2}$ to convert from {\em nats} to {\em bits} (equivalently, change the base $\ln$ to $\log_2$ since $\frac{\ln x}{\ln2} = \log_2 x$):
 $H_Q \leq \frac{1}{2\ln 2} - \left(1+2A\right) \log_2 A$.
In terms of the shift function's standard deviation $\sigma$, this gives our result
\begin{align}
H_Q \leq \frac{1}{2\ln 2} - \left(1+4\sqrt{2\pi}\sigma\right) \log_2 2\sqrt{2\pi}\sigma.
\end{align}

In the limit of small $\sigma$, the leading term of the entropy thus scales with $-\log_2 \sigma$,
 such that halving the width of the standard deviation adds one bit to the maximum required quantum memory cost.

\vspace*{0.5em}
\inlineheading{Example 2: Uniform white noise.}
The normalized top-hat (rectangular) shift function allowing for jumps of up to $\pm\Delta$ around a constant displacement $\mu$ is written
\begin{equation}
S_\Delta(x) = \begin{cases}
\frac{1}{2\Delta} & \mathrm{for~} \mu-\Delta \leq x \leq \mu+\Delta, \\
0 & \mathrm{otherwise}.
\end{cases}
\end{equation}

\inlinesubheading{Derivation of eigenvalues.}
Taking the square root of this function alters its normalization, but not its shape:
$\sqrt{S_\Delta(x)} = \sqrt{2\Delta} S_\Delta(x)$.

Suppose $0 < \Delta < \frac{1}{2}$.
In this case, $S_\Delta(x) \conv S_\Delta(-x)$ yields the triangle function
\begin{equation}
S_\Delta(x) \conv S_\Delta(-x) = \begin{cases}
1- \frac{x}{2\Delta} & \mathrm{for~} 0 \leq x \leq 2 \Delta, \\
1 + \frac{x}{2\Delta} & \mathrm{for~} -2\Delta \leq x < 0 , \\
0 & \mathrm{otherwise}.
\end{cases}
\end{equation}
This function is independent of the constant displacement $\mu$.
Indeed, non-zero $\mu$ only results in perfectly cancelling terms $e^{2\pi i k \mu}$ and $e^{-2\pi i k \mu}$ in the Fourier transform.

Basic Fourier analysis tells us that $S_\Delta(x)$ transforms into a {\em normalized sinc function} ($\sinc x = \sin(\pi x) / \pi x$), and the triangle function into the square of this:
$\Fo{S_\Delta(x) \conv S_\Delta(-x)} = 2 \Delta \mathrm{sinc}^{2}(2 k \Delta)$.
As this tends to $0$ for large $k$, we can approximate the values of $\lambda_k$ for large $N$ using eq.~\eqref{eq:AsymEigen}, to find the eigenspectrum
\begin{equation}
\label{eq:SquareTransApp}
\lambda_k = 2\Delta  \mathrm{sinc}^2\!\left(2 k \Delta \right) \quad \mathrm{for~} k = -\frac{N}{2}, \ldots \frac{N}{2}-1.
\end{equation}

\inlinesubheading{Upper bound on quantum memory cost.}
Through the careful deployment of mildly intimidating algebra,
 we can also derive an upper bound on entropy cost of simulating the square shift function.
The outline of the proof is as follows.
To bound $\sum_k c(k)$ where $c(k) = -\lambda_k \ln \lambda_k$,
 we first construct a monotonically decreasing function $d(k)$ that satisfies $c(k)\leq d(k)$ at every $k$,
 and then show that $\sum d(k)$ is bounded from above.
This sum will hence also upper-bound $\sum_k c(k)$.
As with the Gaussian example, for algebraic convenience, we will use natural logarithms and only consider the region of positive $k$.
In the final stage, we will convert from {\em nats} to {\em bits}, and use the evenness of $c(k)$ to arrive at the full bound.

Explictly, we write
\begin{equation}
c(x) = -2 \Delta \dfrac{\sin^2 x}{x^2} \ln \left( 2\Delta \dfrac{\sin^2 x}{x^2} \right)
\end{equation}
where we have made the substitution $x=2\pi k \Delta$.

In the region $x>0$, we can expand
\begin{equation}
c(x) = -2 \Delta \dfrac{\sin^2 x}{x^2} \left[\ln \left( {\sin^2 x} \right) - 2 \ln \left( \frac{x}{\sqrt{2\Delta}} \right)\right].
\end{equation}
The function $-y \ln y$ has a maximum value of $\frac{1}{e}$ at $y=e$,
 and so we can upper bound $c(x)$ by making the substitution of $- \sin^2 x \ln \sin^2 x$ with $\frac{1}{e}$. 
Since $\sin^2\!x \in [0,1]$,
 in the region $x>\sqrt{2\Delta}$ where $4 \Delta \ln \left(\frac{x}{\sqrt{2\Delta}}\right) > 0$,
 we can likewise upper bound $c(x)$ by making the substitution of $\sin^2\!\left( x \right)$ with $1$.
Thus, for the region $x>\sqrt{2\Delta}$, we have a function $f(x) \geq c(x)$ given
\begin{equation}
f(x) = \dfrac{2 \Delta}{x^2} \left[ \frac{1}{e} + 2 \ln \frac{x}{\sqrt{2\Delta}} \right].
\end{equation}

However, as we plan to ultimately apply the Maclaurin--Cauchy integral convergence test, it is only convenient to use this upper bound in the region of $x$ where $f(x)$ monotonically decreases.
We identify this region by setting $\frac{df}{dx}=\frac{4\Delta}{x^3}\left(-2\ln \frac{x}{\sqrt{2\Delta}} + 1 - \frac{1}{e}\right)=0$,
 to find that $f(x)$ decreases monotonically
 when $x \geq \sqrt{2\Delta} \exp\left(\frac{e-1}{2e}\right)$,
 descending from its maximum value of $\exp\left(\frac{1-e}{e}\right)$.

However, once again consider $c(x)$.
Since it has the form of $-y\ln y$, it follows that in {\em any} region, $c(x) \leq \frac{1}{e}$.
Since $\frac{1-e}{e}>\frac{1}{e}$,
 we can then upper bound $c(x)$ in the region of $0 \leq x \leq \sqrt{2\Delta}$ to form the monotonically decreasing function $d(x)$ given
\begin{equation}
d(x) = \begin{cases}
\exp\left(\frac{1-e}{e}\right)
 & \quad 0 \leq x \leq \sqrt{2\Delta} \exp\left(\frac{e-1}{2e}\right) \\
 \dfrac{2 \Delta}{x^2} \left[ \frac{1}{e} + 2 \ln \frac{x}{\sqrt{2\Delta}} \right]
 & \quad x >  \sqrt{2\Delta}  \exp\left(\frac{e-1}{2e}\right),
\end{cases}
\end{equation}
that is guaranteed to satisfy $d(x) \geq c(x)$ for all $x\geq 0$.
At this point, it is convenient to express this again in terms of $k$,
 making the substitution $k_{\rm split} = \frac{1}{\pi \sqrt{2\Delta}} \exp\left(\frac{e-1}{2e}\right)$:
\begin{equation}
d(k) = \begin{cases}
\exp\left(\frac{1-e}{e}\right)
 & 0 \leq k \leq \lceil k_{\rm split} \rceil \\
\dfrac{1}{2 \Delta \pi^2 \; k^2} \left[ \frac{1}{e} + 2 \ln \left(\pi \sqrt{2\Delta}  k\right) \right]
 & k > \lceil k_{\rm split} \rceil,
\end{cases}
\end{equation}
where $\lceil k_{\rm split} \rceil$ represents the lowest integer above (or including) $k_{\rm split}$.
This rounding is necessary since $k_{\rm split}=\frac{1}{\pi\sqrt{2\Delta}} \exp\left(\frac{e-1}{2e}\right)$ is in general not an integer.
To upper bound $c(k)$ at all points, we must round up this split between the regions of $k$,
 since $\exp\left(\frac{1-e}{e}\right)$ upper bounds all $f(k)$.
(I.e.\ being slightly too inclusive in the first region will result in a slightly higher value of $d(k)$ for the first $k$ satisfying $k\geq k_{\rm split}$).

Having derived our monotonically decreasing function $d(k)$, we are now in a position to show that $\sum_{k=0}^\infty d(k)$ is finite for $\Delta > 0$.
Writing
$\sum_{k=0}^{\lceil k_{\rm split} \rceil} d(k) + \sum_{k=\lceil k_{\rm split} \rceil}^\infty d(k)$
(for an upper bound, it is fine if a term is counted twice!),
 we evaluate the two regions separately.
Firstly,
\begin{align}
\sum_{k=0}^{\lceil k_{\rm split} \rceil} d(k) & \leq
\left(1+\frac{1}{\pi \sqrt{2\Delta}} \exp\left(\frac{e-1}{2e}\right)\right) \exp\left(\frac{1-e}{e}\right) \nonumber \\
& = \exp\left(\dfrac{1-e}{e}\right) + \dfrac{1}{\pi\sqrt{2\Delta}}\exp\left(\dfrac{1-e}{2e}\right)
\end{align}
 where we have used $\frac{1}{2\pi\Delta} \exp\left(\frac{e-1}{2e}\right) +1 > \lceil \frac{1}{2\pi\Delta} \exp\left(\frac{e-1}{2e}\right) \rceil $.
Secondly, using the Maclaurin-Cauchy integral test (see e.g.~\cite{Knopp90}), we bound
\begin{align}
\sum_{k=\lceil k_{\rm split} \rceil}^\infty d(k) & \leq d(\lceil k_{\rm split} \rceil) + \int_{\lceil {k_{\rm split}} \rceil}^\infty d(k) dk \nonumber \\
& \leq  \exp\left(\frac{1-e}{e}\right)+ \int_{k_{\rm split}}^\infty d(k) dk,
\end{align}
where the second line follows by substituting $d(\lceil k_{\rm split} \rceil)$ with the maximum value of $d(k)$,
 and by failing to round up the lower bound of the integral (thus including an extra contribution equal to $\int_{k_{\rm split}}^{\lceil k_{\rm split} \rceil} d(k) dk \geq 0$).
This integral may be analytically solved,
\begin{align}
\dfrac{1}{2 \pi^2 \Delta } \int_{k_{\rm split}}^\infty
  \frac{1}{k^2} \left[ \frac{1}{e} + 2 \ln \left(\pi\sqrt{2 \Delta} k\right) \right] \hspace{-14.5em} & \nonumber \\
& = \left[ \dfrac{-1}{2 \pi^2 \Delta \; k} \left( \frac{1}{e} + 2 + 2\ln \left(\pi\sqrt{2 \Delta} k\right) \right) \right]_{\frac{1}{\pi\sqrt{2\Delta}}\exp\left(\frac{e-1}{2e}\right)}^\infty \nonumber \\
& = \frac{3}{\pi\sqrt{2\Delta}} \exp\left(\dfrac{1-e}{2e}\right).
\end{align}

Combining these two terms, we arive at:
\begin{equation}
\label{eq:SemiBound}
\sum_{k=0}^{\infty} d(k) \leq \frac{4}{\pi\sqrt{2\Delta}} \exp\left(\dfrac{1-e}{2e}\right) + 2 \exp\left(\dfrac{1-e}{e}\right).
\end{equation}

Finally, to bound the entropy $H_Q=-\sum_{\infty}^{\infty} \lambda_k \log_2 \lambda_k$, we must double the above ($c(k)$ is even, and equation~\eqref{eq:SemiBound} bounds only the region $[0,\infty)$),
 and we convert from nats to bits (by including a factor of $\frac{1}{\ln 2}$):
\begin{equation}
H_{Q} \leq \frac{8}{\pi\ln 2\sqrt{2\Delta}} \exp\left(\dfrac{1-e}{2e}\right) + \frac{4}{\ln 2} \exp\left(\dfrac{1-e}{e}\right).
\end{equation}

By evaluating the constant terms, approximately,
\begin{equation}
H_{Q} \leq \frac{1.894}{\sqrt{\Delta}} + 3.067,
\end{equation}
yielding our result.


\end{document}